\documentclass[twocolumn,aps,prd,nofootinbib]{revtex4-1}

\pdfoutput=1

\usepackage[natural,dvipsnames,hyperref,table]{xcolor}
\usepackage{graphicx}

\usepackage[T1]{fontenc}
\usepackage[ansinew]{inputenc}
\usepackage[english]{babel}
\newcommand\ie{{i.e.}~}
\newcommand\eg{{e.g.}~}

\usepackage[
	colorlinks=true,
	linkcolor=Blue,
	citecolor=Blue,
	urlcolor=Blue
]{hyperref}

\newcommand{\DOI}[1]{[\href{http://dx.doi.org/#1}{\tt url}]}
\newcommand\arXiv[2]{[\href{http://arxiv.org/abs/#1}{\tt arXiv:#1}] [#2]}
\newcommand\arXivold[1]{[\href{http://arxiv.org/abs/#1}{\tt #1}]}

\usepackage{amsmath,amssymb,amsfonts}
\usepackage{upgreek}  
\usepackage{suffix}   

\usepackage{cleveref} 
\crefname{equation}{Eq.}{Eqs.}
\crefname{section}{Section}{Sections}
\crefname{figure}{Fig.}{Figs.}
\crefname{table}{Table}{Tables}

\allowdisplaybreaks

\newcommand\nn{\nonumber}

\newcommand\eq[1]{
	\begin{align}
	#1
	\end{align}
}
\WithSuffix\newcommand\eq*[1]{
	\begin{align*}
	#1
	\end{align*}
}
\newcommand\subeq[1]{
	\begin{subequations}
	\eq{#1}
	\end{subequations}
}
\WithSuffix\newcommand\subeq*[1]{
	\begin{subequations}
	\eq*{#1}
	\end{subequations}
}

\newcommand\weq[1]{
	\begin{widetext}
	\eq{#1}
	\end{widetext}
}
\WithSuffix\newcommand\weq*[1]{
	\begin{widetext}
	\begin{align*}
	#1
	\end{align*}
	\end{widetext}
}
\newcommand\wsubeq[1]{
	\begin{widetext}
	\begin{subequations}
	\eq{#1}
	\end{subequations}
	\end{widetext}
}
\WithSuffix\newcommand\wsubeq*[1]{
	\begin{widetext}
	\begin{subequations}
	\eq*{#1}
	\end{subequations}
	\end{widetext}
}

\newcommand\of[1]{\left({#1}\right)}

\newcommand\vev[1]{\langle{#1}\rangle}

\newcommand\is[1]{\stackrel{#1}{=}}

\newcommand\mustbe{\is{!}}

\newcommand\re{{\rm Re}}

\newcommand\tr{{\rm Tr}}

\newcommand\half{\frac{1}{2}}
\newcommand\der[2]{\frac{d{#2}}{d{#1}}}
\newcommand\pder[2]{\frac{\partial{#2}}{\partial{#1}}}
\newcommand\at[2]{\left.{#1}\right\vert_{#2}}

\newcommand\D{{\cal D}}
\newcommand\Nc{N_{\rm c}}
\newcommand\q{\psi}
\newcommand\qbar{\overline{\q}}
\newcommand\qqbar[1]{\qbar_{#1}\q_{#1}^{}}
\newcommand\Dq{\D\q\D\qbar}

\usepackage{ulem}

\begin{document}

\title{Strong-Coupling Lattice QCD on Anisotropic Lattices}

\author{Philippe de Forcrand$^{1,2}$}
\email{forcrand@phys.ethz.ch}
\author{Wolfgang Unger$^{3}$}
\email{wunger@physik.uni-bielefeld.de}
\author{H\'{e}lvio Vairinhos} 
\email{helvio.vairinhos@gmail.com}
\affiliation{$^{1}$Institut f\"{u}r Theoretische Physik, ETH Z\"{u}rich, CH-8093 Z\"{u}rich, Switzerland.}
\affiliation{$^{2}$CERN, TH Department, CH-1211 Geneva 23, Switzerland.}
\affiliation{$^{3}$Fakult\"{a}t f\"{u}r Physik, Universit\"{a}t Bielefeld, Universit\"{a}tstrasse 25, D33619 Bielefeld, Germany.}

\begin{abstract}
Anisotropic lattice spacings are mandatory to reach the high temperatures where chiral symmetry is restored in the strong coupling limit of lattice QCD. Here, we propose a simple criterion for the nonperturbative renormalisation of the anisotropy coupling in strongly-coupled SU($\Nc$) or U($\Nc$) lattice QCD with massless staggered fermions. We then compute the renormalised anisotropy, and the strong-coupling analogue of Karsch's coefficients (the running anisotropy), for $\Nc=3$. We achieve high precision by combining diagrammatic Monte Carlo and multi-histogram reweighting techniques. We observe that the mean field prediction in the continuous time limit captures the nonperturbative scaling, but receives a large, previously neglected correction on the unit prefactor. Using our nonperturbative prescription in place of the mean field result, we observe large corrections of the same magnitude to the continuous time limit of the static baryon mass, and of the location of the phase boundary associated with chiral symmetry restoration. In particular, the phase boundary, evaluated on different finite lattices, has a dramatically smaller dependence on the lattice time extent. We also estimate, as a byproduct, the pion decay constant and the chiral condensate of massless SU(3) QCD in the strong coupling limit at zero temperature.
\end{abstract}

\maketitle

\section{Introduction}

For all practical purposes, the sign problem in lattice QCD with staggered fermions at finite density has been solved at strong coupling. By integrating out the gauge degrees of freedom exactly -- which allows replacing Grassmann integration by a sum over fermionic colour singlets -- the sign problem becomes mild enough to allow for controlled numerical results at moderate volumes, by combining importance sampling and reweighting methods. As a result, the phase diagram of lattice QCD in the strong coupling limit \cite{deForcrand:2009dh} and at first order in the strong coupling expansion \cite{deForcrand:2014tha} can be completely mapped.

In practice, however, it is not sufficient to simulate the strongly-coupled theory directly on rectangular lattices, because the critical temperature of chiral symmetry restoration is higher than what can be reached using the smallest lattice time extent.\footnote{With staggered fermions, the spacetime lattice is necessarily bipartite. In particular, on a rectangular lattice it has an even number of lattice points in each direction. In this case, the lattice time extent is $N_t \geq 2$, hence the lattice temperature is $aT = \frac{1}{N_t} \leq 0.5 < aT_c$ .} In order to study the thermodynamical properties of staggered lattice QCD, in particular across the chiral phase transition, it is therefore necessary to simulate the theory on anisotropic lattices.

On anisotropic lattices, one assigns independent lattice spacings to the spatial and temporal directions, respectively $a$ and $a_t$. The corresponding physical extents of the lattice can then be varied continuously, and independently. A more useful parameterisation of the lattice geometry uses the spatial lattice spacing, $a$, and the anisotropy parameter $\xi$, 
\eq{
\xi = \frac{a}{a_t}
} 
which becomes unity when the lattice is isotropic, and diverges in the continuous time limit $a_t \to 0$. In this parameterisation, the lattice temperature is given by:
\eq{
aT = \frac{\xi}{N_t}
}
where $N_t$ is the lattice time extent. Hence, the lattice temperature can be varied continuously, through $\xi$.

In lattice gauge theory, the physical parameters $a$ and $\xi$ can only be varied implicitly, through independent bare parameters: the bare gauge coupling $\beta$ and the bare anisotropy coupling $\gamma$.
These bare parameters couple differently to the spatial and temporal plaquettes in the Wilson action of SU($\Nc$) or U($\Nc$) pure lattice gauge theory in $d+1$ dimensions \cite{aniso}:
\weq{
S_g = \frac{\beta}{\gamma} \sum_x \sum_{1\leq i<j\leq d} \of{1 - \frac{1}{\Nc} \re\tr\of{U_{x,ij}}}
+ \beta \gamma \sum_x \sum_{i=1}^d \of{1 - \frac{1}{\Nc} \re\tr\of{U_{x,i0}}}
\label{eq:pure-gauge-action}
}
where $U_{x,\mu\nu}$ is the ordered product of link variables around a plaquette parallel to the $\hat\mu$ and $\hat\nu$ directions.

For a single flavour of staggered fermions in the strong coupling limit ($\beta=0$), the anisotropic lattice action is given by:
\weq{
S_f = 2a_t m_q \sum_x \qqbar{x} + \sum_{x} \sum_{\mu=0}^d \gamma^{\delta_{\mu 0}} \eta_{x\mu} \of{e^{a_t\upmu_q\delta_{\mu 0}}\qbar_x U_{x\mu} \q_{x+\hat\mu} - e^{-a_t\upmu_q\delta_{\mu 0}}\qbar_{x+\hat\mu} U_{x\mu}^\dag \q_x}
\label{eq:staggered-action}
}
where $a_t m_q$ and $a_t \upmu_q$ are the bare quark mass and quark chemical potential, respectively, and $\eta_{x\mu} = \pm 1$ are the staggered phases. In the case of U($\Nc$), gauge invariance dictates that colour singlets are independent of $a_t \upmu_q$, hence we may set $a_t \upmu_q$ to zero without loss of generality.

How $a$ and $\xi$ depend on the bare parameters of the theory is unknown a priori. This knowledge is, however, essential for precision measurements on anisotropic lattices, \eg bulk thermodynamic quantities, and any uncontrolled approximation can easily be the main source of systematic errors. 

In the weak gauge coupling regime ($\beta\to\infty$) of the SU($\Nc$) pure gauge theory \cref{eq:pure-gauge-action}, perturbation theory and the non-renormalisation of the speed of light can be used to calibrate the anisotropy coupling \cite{aniso-pert}. In that regime, it is found that $\xi_{\rm pert}(\gamma) = \gamma$ (as expected classically).

Using mean field techniques, the behaviour of the renormalised anisotropy at strong coupling ($\beta \ll 1$) and at large values of $\gamma$ is predicted to be quadratic, with unit prefactor \cite{aniso-mf}:
\eq{
\xi_{\rm mf}(\gamma) = \gamma^2
\label{eq:mf}
}

In the nonperturbative regime, however, the relation between bare and renormalised anisotropy couplings can only be determined numerically. This has been done, for example, in pure gauge theory \cite{aniso,aniso-gauge}, in lattice QCD with staggered fermions \cite{aniso-stagg} or Wilson fermions \cite{aniso-wilson}. The nonperturbative renormalisation of the bare parameters requires fine-tuning, guided by some physical criterion which controls the recovery of Euclidean symmetry.

In this Letter we present a simple, precise, and nonperturbative method to calibrate the anisotropy coupling in lattice QCD with massless staggered fermions, in the limit of strong gauge coupling.

\section{Diagrammatic representation of lattice QCD}

The partition function of SU($\Nc$) or U($\Nc$) QCD on a bipartite $N_t \times N_s^d$ lattice, with a single flavor of staggered fermions, in the strong coupling limit ($\beta\to 0$) factorises into a product of solvable fermionic one-link integrals:
\weq{
Z = \int \Dq \, \exp\of{2a_t m_q \sum_x \qqbar{x}} \prod_{x,\mu} \int dU_{x\mu} \, 
\exp\of{\gamma^{\delta_{\mu 0}} \eta_{x\mu} \of{e^{a_t\upmu_q\delta_{\mu 0}}\qbar_x U_{x\mu} \q_{x+\hat\mu} - e^{-a_t\upmu_q\delta_{\mu 0}}\qbar_{x+\hat\mu} U_{x\mu}^\dag \q_x}}
}

In the SU($\Nc$) case, the group integration of the link variables, followed by the Grassmann integration of the fermionic degrees of freedom, yields the partition function of a monomer-dimer-loop system \cite{mdp}:
\weq{
Z = \sum_{\{n,k,\ell\}} 
{\prod_x \frac{\Nc!}{n_x!}}\,
{\prod_{x,\mu} \frac{(\Nc - k_{x\mu})!}{\Nc! k_{x\mu}!}}
\frac{\sigma(\ell)}{\Nc!^{\vert \ell \vert}} \,
(2a_t m_q)^{N_M} \,
\gamma^{2N_{Dt} + \Nc N_{\ell t}} \,
e^{\Nc N_t a_t \upmu_q w(\ell)}
\label{eq:partition-mdp}
}
This partition function is a constrained sum over integer occupation numbers of monomers and dimers, $n_x, k_{x\mu} \in \{0,1,\ldots,\Nc\}$, and of oriented baryon links, $\ell_{x\mu}\in\{0,\pm 1\}$, which combine to form oriented baryon loops. The global quantities:
\subeq{
N_{M} = \sum_x n_x, 
\\
N_{Dt} = \sum_x k_{x0}, 
\\
N_{\ell t} = \sum_x \vert \ell_{x0} \vert
}
enumerate the monomers, temporal dimers, and temporal baryon links on the lattice, respectively. $\sigma(\ell) = \pm 1$ is a geometric sign associated with the configuration of baryon loops $\ell$, $\vert\ell\vert$ is their length, and $w(\ell)$ is their winding number around the Euclidean time direction.

The monomers represent fermion condensates, $M_x^{n_x}$, dimers represent meson hoppings, $(M_x M_{x+\hat\mu})^{k_{x\mu}}$, and baryon links represent baryon hoppings, $\overline{B}_x B_{x+\hat\mu}$ or $\overline{B}_{x+\hat\mu} B_x$, where $M_x$ is a meson and $B_x$ is a baryon:
\subeq{
M_x &= \qbar_x\q_x^{}
\\
B_x &= \frac{1}{\Nc!} \varepsilon_{i_1 \cdots i_{\Nc}}\q_x^{i_1} \cdots \q_x^{i_{\Nc}}
}

In order for a configuration of occupation numbers to contribute non-trivially to the partition function \cref{eq:partition-mdp}, the Grassmann integrals over the corresponding fermionic degrees of freedom must be non-trivial on each lattice site. 

Due to their Grassmann nature, such configurations must necessarily represent arrangements of exactly $\Nc$ fermions and $\Nc$ anti-fermions on each lattice site.\footnote{If the gauge group is SU($\Nc$), the ordering of the Grassmann variables in such arrangements contributes with the geometric sign $\sigma(\ell) = \pm 1$, which introduces a (baryonic) sign problem in the system. See \cref{eq:partition-mdp}.} This imposes the following local constraints on the integer occupation numbers:
\subeq{
n_x + \sum_{\pm\mu} \of{k_{x\mu} + \frac{N_c}{2}\vert\ell_{x\mu}\vert} &\mustbe N_c
\label{eq:md-constraint}
\\
\sum_{\pm\mu} \ell_{x\mu} &\mustbe 0
\label{eq:l-constraint}
}
\cref{eq:l-constraint} is a local discrete conservation law for baryon links, which formalises our statement above that baryon links in admissible configurations form closed oriented loops.

In the U($\Nc$) case, since $\ell_{x\mu} = 0$, the partition function \cref{eq:partition-mdp} reduces to a sum over monomer-dimer configurations:
\eq{
Z = \sum_{\{n,k\}} 
{\prod_x \frac{\Nc!}{n_x!}}\,
{\prod_{x,\mu} \frac{(\Nc - k_{x\mu})!}{\Nc! k_{x\mu}!}}
(2a_t m_q)^{N_M} 
\gamma^{2N_{Dt}}
\label{eq:partition-mdp-Un}
}
with the same Grassmann constraint for monomers and dimers on each site:
\eq{
\qquad &n_x + \sum_{\pm\mu} k_{x\mu} \mustbe \Nc
\label{eq:constraint-Un}
}
Likewise, the U($\Nc$) observables are defined in the same way as the observables (in the mesonic sector) of the SU($\Nc$) theory.

\section{Conserved currents and conserved charges}

Let $\sigma_x = \pm 1$ be the parity of the site $x$ on a bipartite lattice.
From \cref{eq:l-constraint}, it is easy to construct baryonic currents:
\eq{
j^B_{x\mu} = \sigma_x \ell_{x\mu}
\label{eq:baryon-currents}
}
which are conserved at every site:
\eq{
\sum_{\mu=0}^d \of{j^B_{x\mu} - j^B_{x-\hat\mu,\mu}} = 0
}
The corresponding conserved charges are integrals of the baryonic currents \cref{eq:baryon-currents} over a codimension-1 lattice slice ${\cal S}_\mu$, perpendicular to $\hat\mu$:
\eq{
Q^B_\mu = \sum_{x\in{\cal S}_\mu} j^B_{x\mu}
}

Similarly, by rewriting \cref{eq:md-constraint} as:
\eq{
\sum_{\pm\mu} \of{k_{x\mu} + \frac{\Nc}{2} \vert \ell_{x\mu} \vert - \frac{\Nc}{2d}} = -n_x
}
it is easy to construct the corresponding (pion) currents:
\eq{
j_{x\mu} = \sigma_x \of{k_{x\mu} + \frac{\Nc}{2} \vert \ell_{x\mu} \vert - \frac{\Nc}{2d}}
\label{eq:pion-current}
}
from which a local discrete Gauss' law for dimers results:
\eq{
\sum_{\mu=0}^d \of{j_{x\mu} - j_{x-\hat\mu,\mu}} = -\sigma_x n_x
\label{eq:pion-current-conservation}
}

Thus, monomers are sources of the pion currents. Using Grassmann variables, the source term on the r.h.s. of \cref{eq:pion-current-conservation} corresponds to $-a_t m_q\qbar_x \gamma_5 \q_x$. Only in the chiral limit, \ie in the absence of monomers, are the pion currents conserved. 
In the chiral limit, the corresponding conserved charges are integrals of the pion currents over a lattice slice ${\cal S}_\mu$:
\eq{
Q_\mu = \sum_{x \in {\cal S}_\mu} j_{x\mu}
\label{eq:pion-charge}
}

In the U($\Nc$) theory, since $\ell_{x\mu} = 0$, the pion currents simplify to:
\eq{
j_{x\mu} = \sigma_x \of{k_{x\mu} - \frac{\Nc}{2d}}
\label{eq:pion-current-Un}
}

\section{\label{sect:calibration}Nonperturbative anisotropy calibration}

In this Section, we show how the conserved pion charges can be used to calibrate the anisotropy coupling in lattice QCD with staggered fermions, at zero temperature, in the strong coupling limit.

In the strong coupling limit, the partition functions of SU($\Nc$) and U($\Nc$) lattice QCD with staggered fermions have monomer-dimer-loop representations, \cref{eq:partition-mdp,eq:partition-mdp-Un}, with no dependence on the spatial lattice spacing, $a$. In order for the pion charges $Q_\mu$ to be conserved, we take the lattice fermions to be massless, $a_t m_q = 0$. In the SU($\Nc$) case, we only consider the case of zero chemical potential, $a_t \upmu_q = 0$.\footnote{The chemical potential only modifies the temporal boundary conditions, which is irrelevant at $T=0$. A non-zero quark mass, on the other hand, modifies the dynamics, and so the renormalization prescription must take this into account (we discuss the massive case in the Conclusion).} The corresponding partition functions thus depend only on a single parameter: the bare anisotropy coupling $\gamma$.

Let us consider the theories to be defined on anisotropic $N_t \times N_s^d$ lattices. In order to calibrate the anisotropy, we compare the fluctuations of the conserved pion charges in different directions. 

Due to spatial isotropy, the expectation values of fluctuations of the spatial pion charges $Q_i, i=1,\ldots,d$ must coincide. Therefore, it is convenient to quantify spatial fluctuations using the expectation value of:
\eq{
Q_s^2 = \frac{1}{d} \sum_{i=1}^d Q_i^2
\label{eq:spatial-charge-fluctuation}
}
while the temporal fluctuations are quantified using the expectation value of $Q_t^2 = Q_0^2$.

Now, when the lattice is hypercubic, \ie $N_t = \xi N_s$, the fluctuations of the spatial and temporal conserved charges must be equal. This provides a simple, nonperturbative criterion for the renormalisation of the anisotropy coupling: the value of the bare parameter, $\gamma_{\rm np}$, corresponding to the renormalised value, $\xi(\gamma_{\rm np}) = N_t/N_s$, is that for which the fluctuations of the spatial and temporal conserved charges are equal:
\eq{
\vev{Q_t^2}_{\gamma_{\rm np}} \mustbe \vev{Q_s^2}_{\gamma_{\rm np}}
\label{eq:renormalization-criterion}
}

\begin{figure}[t]
\centering
\includegraphics[scale=0.33]{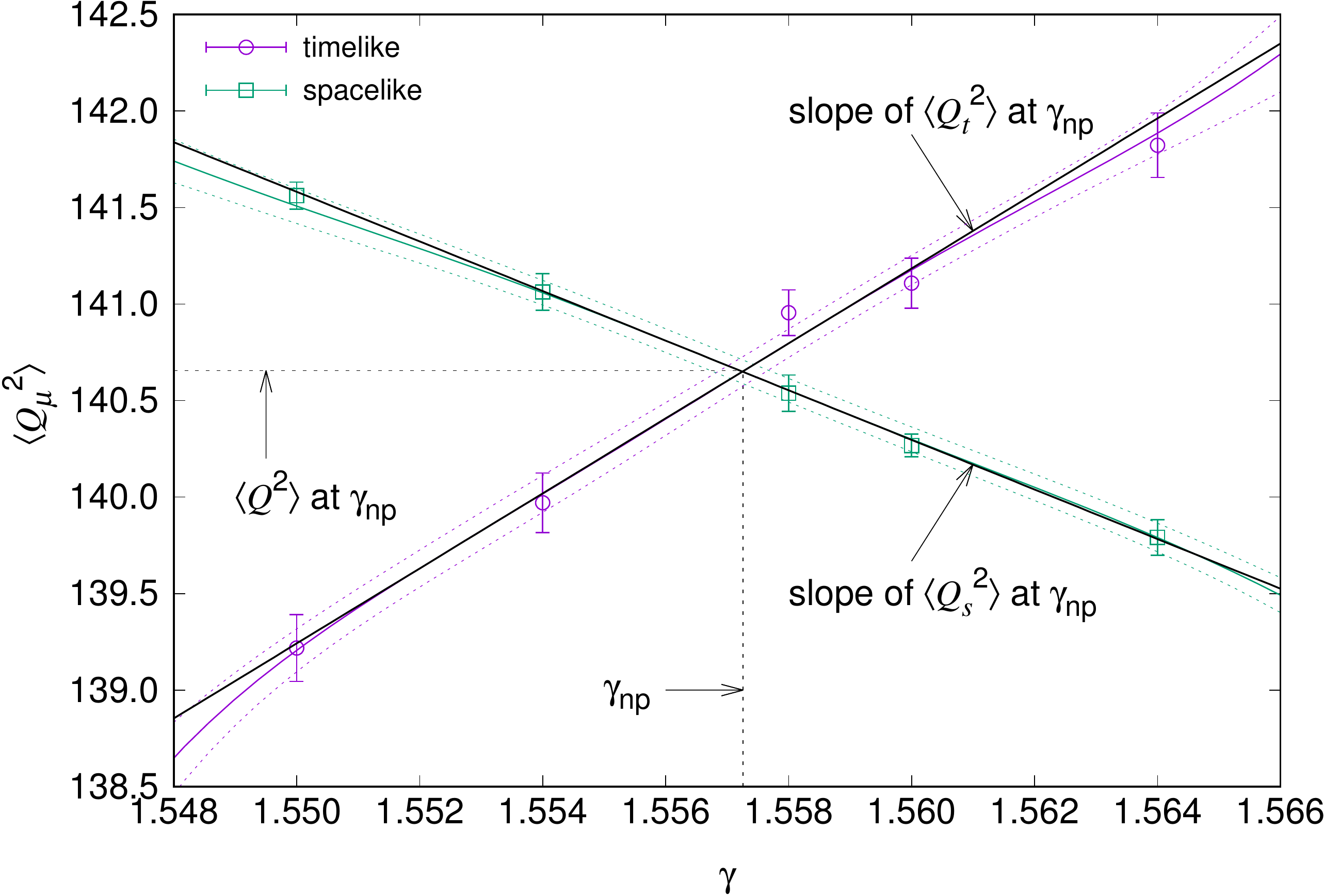}
\caption{\label{fig:intersection}
Measurements of the fluctuations of the conserved pion charges in a numerical simulation of U(3) lattice QCD on a $32 \times 16^3$ lattice. The measurements are interpolated using Ferrenberg-Swendsen multi-histogram reweighting. The intersection of the two curves provides a precise nonperturbative estimate of the bare parameter $\gamma_{\rm np}$ associated with the renormalised anisotropy: $\xi=2$. It also provides an estimate of the value of such fluctuations in the hypercubic lattice, $\vev{Q^2}$, which, together with the estimates of the slopes of the tangents to the curves at the intersection point, allows an estimation of the running anisotropy, $\frac{1}{\xi}\der{\gamma}{\xi}$.}
\end{figure}

In \cref{fig:intersection}, we give a practical example. In a numerical simulation of U(3) lattice QCD on a $32 \times 16^3$ lattice, we evaluate $\vev{Q_s^2}$ and $\vev{Q_t^2}$ for a few values of the bare parameter $\gamma$, about the correct nonperturbative value $\gamma_{\rm np}$ associated with the renormalised anisotropy parameter, $\xi = 2$. Using Ferrenberg-Swendsen multi-histogram reweighting, we interpolate the measurements of the fluctuations, and estimate with high precision the value of the bare parameter for which the two curves intersect, \ie when the lattice is hypercubic. In this particular case, $\gamma_{\rm np} = 1.55725(29)$. This value is to be compared with the commonly accepted mean field prediction, $\gamma_{\rm mf} = \sqrt{\xi=2} \approx 1.41421$.

\section{Running anisotropy}

It is also possible to estimate the running of the anisotropy parameter, $\frac{1}{\xi}\der{\gamma}{\xi}$, using extra information from the intersection point in \cref{fig:intersection}. This quantity -- the strong-coupling analogue of Karsch's coefficients \cite{aniso-pert} -- is important for computing \eg bulk thermodynamic quantities, like the energy density and pressure \cite{deForcrand:2017obe}.

The fluctuations of the conserved charges scale with the volume of the lattice slices on which the corresponding conserved currents are integrated over:
\subeq{
\label{eq:charge-scaling-t}
\vev{Q_t^2} &\propto (N_s a)^3 \\
\label{eq:charge-scaling-s}
\vev{Q_s^2} &\propto (N_s a)^2 N_t a_t
}
The ratio of temporal and spatial fluctuations then becomes directly related to the renormalised anisotropy:
\eq{
\frac{\vev{Q_t^2}}{\vev{Q_s^2}}
= \frac{N_s}{N_t} \xi
\label{eq:charge-variance-ratio}
}
We have already explained the fact that this ratio is 1 when the lattice is hypercubic.

Now, taking the derivative of \cref{eq:charge-variance-ratio} with respect to the bare parameter $\gamma$, at the intersection of the curves in \cref{fig:intersection}, yields the value of the running anisotropy at that point:
\eq{
\at{\der{\gamma}{} \frac{\vev{Q_t^2}}{\vev{Q_s^2}}}{\gamma_{\rm np}}
&= \frac{\vev{Q_t^2}'_{\gamma_{\rm np}} - \vev{Q_s^2}'_{\gamma_{\rm np}}}{\vev{Q^2}_{\gamma_{\rm np}}}
\nn\\
&= \frac{N_s}{N_t} \at{\der{\gamma}{\xi}}{\gamma_{\rm np}}
= \frac{1}{\xi} \at{\der{\gamma}{\xi}}{\gamma_{\rm np}}
\label{eq:running-anisotropy}
}
Therefore, in order to estimate the value of the running anisotropy at $\gamma_{\rm np}$, we also need the value of the fluctuation of the conserved pion charges on a hypercubic lattice:
\eq{
\vev{Q^2}_{\gamma_{\rm np}} \equiv \vev{Q_t^2}_{\gamma_{\rm np}} \mustbe \vev{Q_s^2}_{\gamma_{\rm np}}
}
and the values of the slopes of the tangents to the curves at the intersection point: $\vev{Q_t^2}'_{\gamma_{\rm np}}$ and $\vev{Q_s^2}'_{\gamma_{\rm np}}$.

\section{Numerical renormalisation}

The Monte Carlo sampling of the U($\Nc$) partition function \cref{eq:partition-mdp-Un} is highly efficient when using directed path algorithms \cite{worm}. In the SU($\Nc$) case, observables must be reweighted because of the occurrence of negative-weight baryonic configurations, even at zero chemical potential. However, this sign problem is mild and controllable for moderate lattice volumes \cite{deForcrand:2009dh,deForcrand:2014tha,deForcrand:2017obe}.

We simulate massless U(3) and SU(3) lattice QCD in the strong coupling limit, using the directed path algorithm \cite{worm}, for several values of the renormalised anisotropy $\xi$. For each $\xi$, we estimate the corresponding value of the bare parameter $\gamma_{\rm np}$ on a $(\xi N_s) \times N_s^3$ lattice, for several values of $N_s$, using the method described in \cref{sect:calibration}. We also measure the running of the anisotropy coupling \cref{eq:running-anisotropy}. The results for U(3) and SU(3) are summarised in \cref{tab:gamma-u3,tab:gamma-su3}, respectively.
\begin{figure*}[t]
\centering
\includegraphics[scale=0.33]{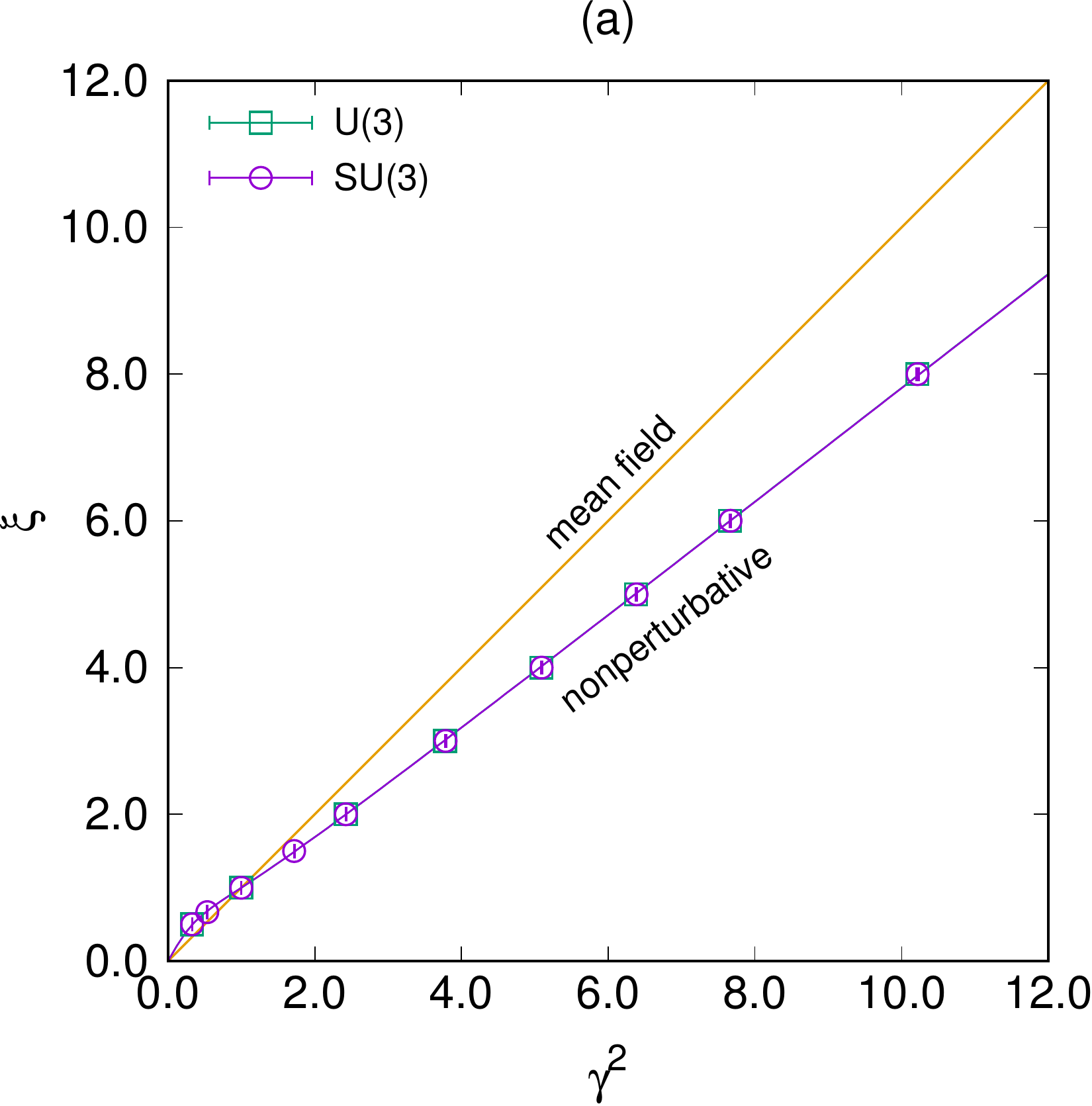}
\qquad
\includegraphics[scale=0.33]{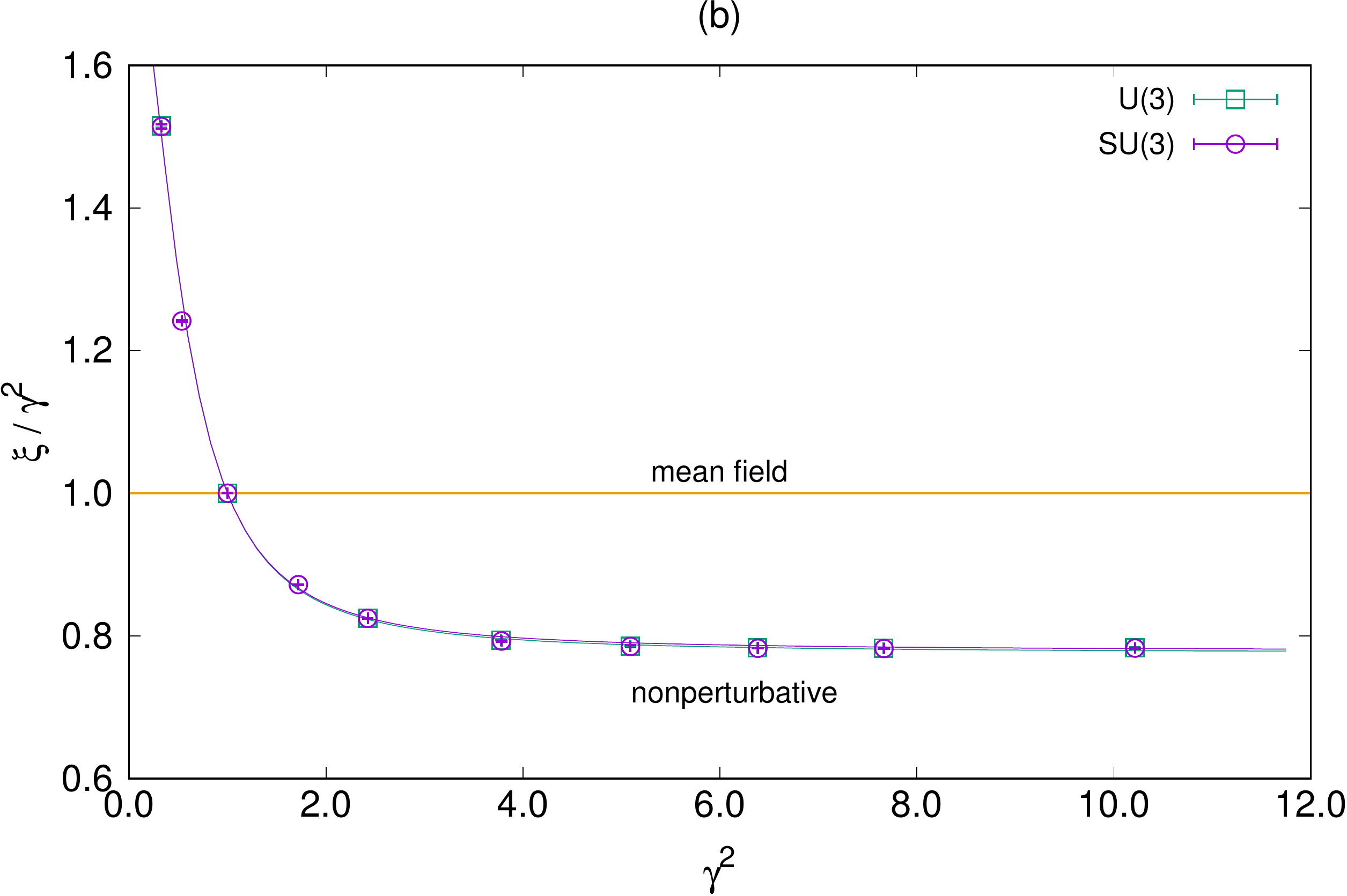}
\includegraphics[scale=0.33]{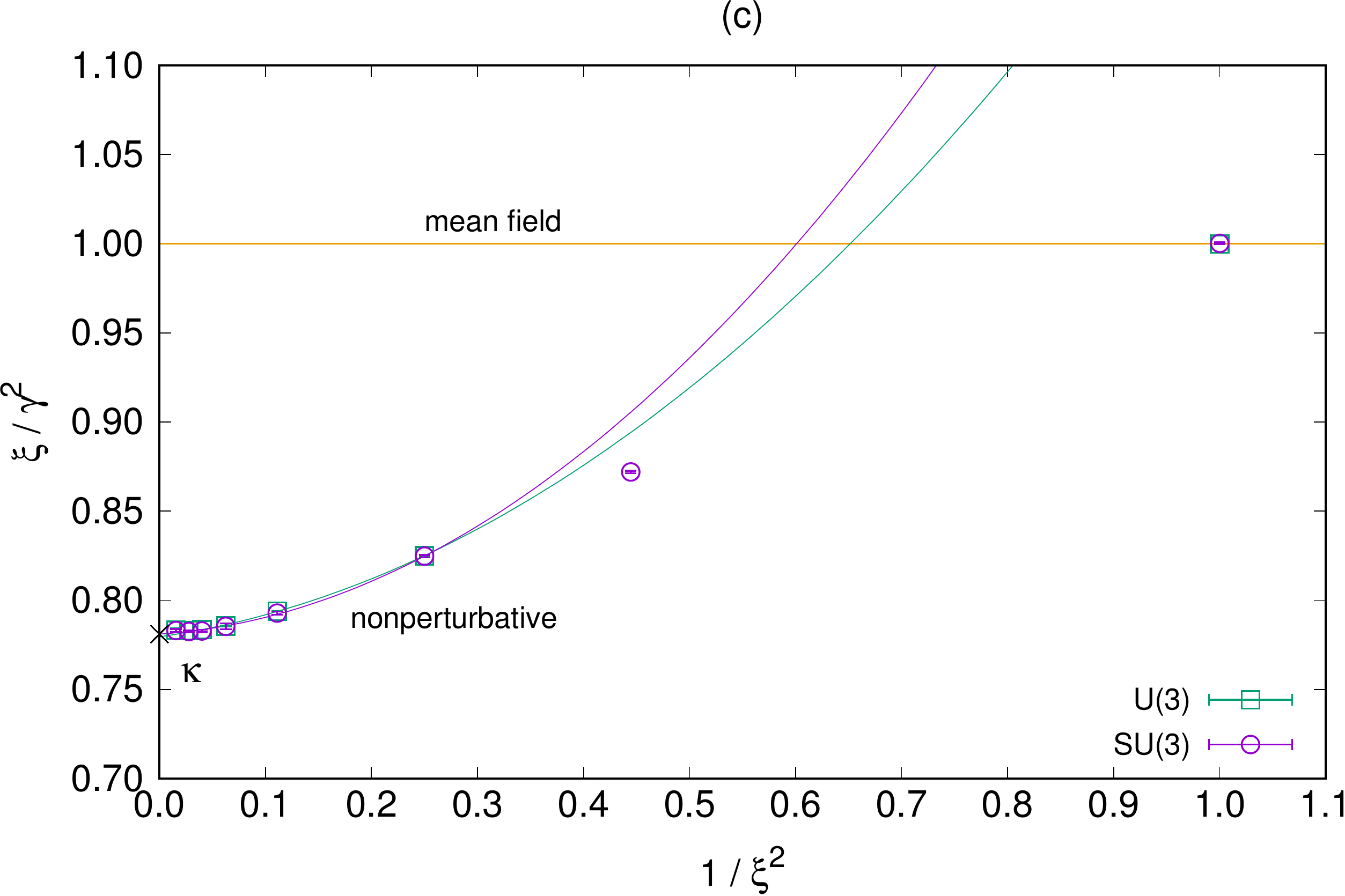}
\caption{
Nonperturbative relation between the bare and renormalised anisotropy parameters, for U(3) (green) and SU(3) (purple) massless lattice QCD, in the thermodynamic limit, presented in 3 different ways. \cref{fig:gamma-np}a shows that, as predicted by mean-field, the renormalised anisotropy at large $\gamma$ is $\xi(\gamma) \propto \gamma^2$, but with a smaller prefactor than predicted. \cref{fig:gamma-np}b shows the ratio $\xi/\gamma^2$ for a wide range of $\gamma$, larger and smaller than 1. A simple one-parameter Ansatz \cref{eq:np} describes the data well. \cref{fig:gamma-np}c shows the approach to the continuous time limit, i.e. $\xi\to\infty$. In that regime, $\xi/\gamma^2$ approaches a constant $\kappa$, with quadratic corrections in $1/\xi^2$. The behaviours of U(3) and SU(3) are almost undistinguishable, because baryons are heavy and describe small loops only.\label{fig:gamma-np}}
\end{figure*}

\begin{table}[t]
\centering
\begin{scriptsize}
\begin{tabular}[t]{llllll}
\hline\hline
$\xi$
& $N_s$
& $\gamma_{\rm np}$
& $\at{\frac{\xi}{\gamma}\der{\xi}{\gamma}}{\gamma_{\rm np}}$
& $a^2 \Upsilon$ 
& $a^6 \chi/N_s^4$
\\
\hline
1/2
& 8  & 0.5741(2)  & 0.435(9) & 0.27470(7)  & 0.283789(5) \\
& 12 & 0.5745(2)  & 0.453(8) & 0.27491(2)  & 0.282628(8) \\
& 16 & 0.5743(2)  & 0.43(1)  & 0.274795(8) & 0.282207(4) \\
& 20 & 0.5743(4)  & 0.44(2)  & 0.27479(2)  & 0.282092(3) \\
& 24 & 0.5744(5)  & 0.44(3)  & 0.27469(2)  & 0.282157(7) \\
\hline
1 
& 4  & 1.00000(5) & 0.357(2) & 0.433247(7) & 0.489464(1) \\
& 6  & 1.0001(5)  & 0.39(2)  & 0.43388(2)  & 0.485824(9) \\
& 8  & 0.9998(5)  & 0.34(2)  & 0.43408(2)  & 0.484272(6) \\
& 10 & 1.0000(5)  & 0.36(3)  & 0.43418(2)  & 0.483406(5) \\
\hline
2 
& 4  & 1.55745(7) & 0.284(2) & 0.548979(9) & 0.683806(2) \\
& 6  & 1.5570(4)  & 0.28(2)  & 0.54933(2)  & 0.67935(1)  \\
& 8  & 1.557(1)   & 0.37(5)  & 0.54945(3)  & 0.67775(2)  \\
& 10 & 1.5565(9)  & 0.27(3)  & 0.54889(3)  & 0.67696(2)  \\
& 12 & 1.5566(8)  & 0.26(3)  & 0.54914(4)  & 0.67636(2)  \\
\hline
3 
& 4  & 1.9446(1)  & 0.261(4) & 0.582224(1) & 0.761084(2) \\
& 6  & 1.9431(8)  & 0.31(2)  & 0.58265(3)  & 0.75674(2)  \\
& 8  & 1.9445(7)  & 0.23(3)  & 0.58247(3)  & 0.75382(2)  \\
& 10 & 1.9442(9)  & 0.25(2)  & 0.58206(4)  & 0.75309(2)  \\
  \hline
4
& 4  & 2.2573(1)  & 0.254(2) & 0.594889(1) & 0.798407(2) \\
& 6  & 2.2566(3)  & 0.257(6) & 0.595057(7) & 0.793426(6) \\
& 8  & 2.2568(4)  & 0.274(6) & 0.59514(2)  & 0.791196(4) \\
& 10 & 2.2566(6)  & 0.268(8) & 0.59497(2)  & 0.79023(1)  \\
  \hline
5 
& 4  & 2.5273(2)  & 0.248(3) & 0.600789(6) & 0.819251(2) \\
& 6  & 2.5267(3)  & 0.264(6) & 0.600829(7) & 0.814061(4) \\
& 8  & 2.5266(5)  & 0.26(2)  & 0.60085(2)  & 0.81195(1)  \\
& 10 & 2.531(3)   & 0.5(3)   & 0.6011(2)   & 0.80865(7)  \\
  \hline
6 
& 4  & 2.7692(2)  & 0.247(2) & 0.603881(5) & 0.832205(3) \\
& 6  & 2.7682(3)  & 0.27(2)  & 0.604074(7) & 0.827064(5) \\
& 8  & 2.7683(6)  & 0.23(2)  & 0.60390(2)  & 0.824761(9) \\
& 10 & 2.7683(8)  & 0.31(2)  & 0.60388(2)  & 0.82384(1)  \\
  \hline
8
& 4  & 3.1954(2) & 0.255(4)  & 0.606741(4) & 0.847192(2) \\
& 6  & 3.1943(5) & 0.238(4)  & 0.60697(2)  & 0.841938(4) \\
& 8  & 3.1946(7) & 0.25(2)   & 0.60665(2)  & 0.83959(2)  \\
& 10 & 3.194(2)  & 0.21(3)   & 0.60687(4)  & 0.83889(1)  \\
\hline\hline
\end{tabular}
\end{scriptsize}
\caption{\label{tab:gamma-u3}
Values of the bare anisotropy coupling $\gamma_{\rm np}$ associated with the renormalised anisotropy $\xi$, from numerical simulations of massless U(3) lattice QCD on $(\xi N_s) \times N_s^3$ lattices. Also, the corresponding values of the running anisotropy (derivative), the helicity modulus $a^2 \Upsilon$, and the chiral susceptibility density $a^6\chi/N_s^4$. The quantity $\gamma_{\rm np}$ exhibits small finite-volume corrections, and is consistent (within errors) with its thermodynamic limit, even on the smallest lattices. This rapid convergence justifies using small-lattice measurements as thermodynamic estimators for $\gamma_{\rm np}$. This is particularly useful in simulations at large $\xi$, for which significant statistics can only be obtained on small volumes.}
\end{table}
\begin{table}[t]
\centering
\begin{scriptsize}
\begin{tabular}[t]{lllllll}
\hline\hline
$\xi$
& $N_s$
& $\gamma_{\rm np}$
& $\at{\frac{\xi}{\gamma}\der{\xi}{\gamma}}{\gamma_{\rm np}}$
& $a^2 \Upsilon$ 
& $a^6 \chi/N_s^4$
& average sign
\\
\hline
1/2
&  8 & 0.5743(2)  & 0.43(1)  & 0.27445(2)  & 0.283424(6) & 0.99657(7)  \\
& 12 & 0.5745(2)  & 0.450(6) & 0.274509(6) & 0.282271(4) & 0.9833(2)   \\
& 16 & 0.5744(2)  & 0.436(6) & 0.274417(6) & 0.281835(2) & 0.9475(8)   \\
& 20 & 0.5744(4)  & 0.43(2)  & 0.274471(6) & 0.281640(4) & 0.818(4)    \\
& 24 & 0.5746(7)  & 0.44(3)  & 0.27459(2)  & 0.28152(2)  & 0.63(2)     \\
\hline
2/3
& 6  & 0.7324(2)  & 0.405(6) & 0.34033(2)  & 0.362517(3) & 0.99863(2)  \\
& 12 & 0.7327(4)  & 0.38(1)  & 0.34040(2)  & 0.359782(8) & 0.9777(4)   \\
\hline
1
&  4 & 0.99993(5) & 0.356(2) & 0.432995(9) & 0.489211(1) & 0.991260(3) \\
&  6 & 1.0000(3)  & 0.36(2)  & 0.43384(2)  & 0.485553(5) & 0.99830(2)  \\
&  8 & 1.0002(3)  & 0.36(2)  & 0.43400(1)  & 0.483803(6) & 0.99543(7)  \\
& 10 & 0.9999(3)  & 0.369(5) & 0.433984(8) & 0.483086(6) & 0.9876(2)   \\
\hline
3/2
&  4 & 1.3117(2)  & 0.309(5) & 0.510010(1) & 0.610195(3) & 0.996258(6) \\
&  8 & 1.3115(5)  & 0.30(2)  & 0.51024(2)  & 0.603968(9) & 0.9933(2)   \\
\hline
2
&  4 & 1.5573(2)  & 0.291(5) & 0.548483(8) & 0.683098(2) & 0.998044(7) \\ 
&  6 & 1.5571(4)  & 0.28(2)  & 0.54882(2)  & 0.678474(8) & 0.99815(3)  \\ 
&  8 & 1.5568(6)  & 0.29(2)  & 0.54884(2)  & 0.676714(8) & 0.99162(2)  \\ 
& 10 & 1.5569(5)  & 0.28(2)  & 0.54873(3)  & 0.67565(2)  & 0.97084(8)  \\ 
& 12 & 1.5572(6)  & 0.24(2)  & 0.54870(2)  & 0.67518(2)  & 0.942(3)    \\
\hline
3
&  4 & 1.9449(2)  & 0.263(4) & 0.581568(5) & 0.760045(5) & 0.999186(4) \\
&  6 & 1.944(1)   & 0.31(6)  & 0.58200(3)  & 0.75514(2)  & 0.99787(8)  \\
&  8 & 1.944(2)   & 0.32(4)  & 0.58200(3)  & 0.75323(2)  & 0.9921(4)   \\
& 10 & 1.945(1)   & 0.26(2)  & 0.58170(4)  & 0.75143(3)  & 0.979(2)    \\
\hline
4
&  4 & 2.2581(6)  & 0.262(8) & 0.59431(2)  & 0.79686(2)  & 0.999682(4) \\
&  6 & 2.2578(9)  & 0.27(2)  & 0.59455(3)  & 0.79164(2)  & 0.99885(5)  \\
&  8 & 2.258(1)   & 0.24(3)  & 0.59433(5)  & 0.78914(4)  & 0.9964(2)   \\
& 10 & 2.2569(6)  & 0.27(1)  & 0.59455(3)  & 0.78899(1)  & 0.9898(5)   \\
\hline
5
&  4 & 2.5288(6)  & 0.25(3)  & 0.6002(2)   & 0.81777(2)  & 0.999770(9) \\
&  6 & 2.527(1)   & 0.21(2)  & 0.60071(4)  & 0.81291(2)  & 0.99885(7)  \\
&  8 & 2.528(2)   & 0.23(3)  & 0.60024(6)  & 0.81009(4)  & 0.9983(2)   \\
& 10 & 2.5272(9)  & 0.25(4)  & 0.60022(4)  & 0.80927(3)  & 0.9870(7)   \\
\hline
6 
&  4 & 2.7702(2) & 0.248(5) & 0.60354(2)  & 0.830977(4) & 0.999816(1) \\
&  6 & 2.7693(4) & 0.241(7) & 0.603662(8) & 0.825741(7) & 0.99958(2)  \\
&  8 & 2.7685(8) & 0.30(3)  & 0.60375(2)  & 0.82393(1)  & 0.99857(6)  \\
& 10 & 2.769(2)  & 0.3(1)   & 0.60362(4)  & 0.82234(5)  & 0.985(1)    \\
\hline
8 
&  4 & 3.1968(3) & 0.257(5) & 0.60656(2)  & 0.845910(3) & 0.999891(1) \\
&  6 & 3.1958(5) & 0.275(8) & 0.60671(2)  & 0.840645(7) & 0.999808(6) \\
&  8 & 3.196(1)  & 0.25(3)  & 0.60669(3)  & 0.83855(2)  & 0.99902(5)  \\
& 10 & 3.195(2)  & 0.22(3)  & 0.60539(7)  & 0.8370(1)   & 0.9935(7)   \\
\hline
\hline
\end{tabular}
\end{scriptsize}
\caption{\label{tab:gamma-su3}
Values of the bare anisotropy coupling $\gamma_{\rm np}$ associated with the renormalised anisotropy $\xi$, from numerical simulations of massless SU(3) lattice QCD on $(\xi N_s) \times N_s^3$ lattices. Also, the corresponding values of the running anisotropy (derivative), the helicity modulus $a^2 \Upsilon$, the chiral susceptibility density $a^6 \chi/N_s^4$, and the average baryonic sign. The quantity $\gamma_{\rm np}$ exhibits small finite-volume corrections, and is consistent (within errors) with its thermodynamic limit, even on the smallest lattices. This rapid convergence justifies using small-lattice measurements as thermodynamic estimators for $\gamma_{\rm np}$. This is particularly useful in simulations at large $\xi$, for which significant statistics can only be obtained on small volumes. In the continuous time limit, the baryons become increasingly static, which explains the lack of fluctuations that contribute to the sign problem at large $\xi$.}
\end{table}

In these tables, rather than storing the estimators of \cref{eq:running-anisotropy}, we instead store the estimators of its reciprocal, the reason being that the latter enters linearly in the definition of important bulk thermodynamic quantities, \eg the energy density:
\eq{
a^4 \varepsilon 
= \frac{a^4}{V} \pder{T^{-1}}{\log Z} 
= \frac{\xi}{\gamma} \der{\xi}{\gamma} \frac{2\xi\vev{N_{Dt}}}{N_s}
}

The nonperturbative relation between the renormalised and bare anisotropy parameters, in the thermodynamic limit, is presented in \cref{fig:gamma-np}a.
At large anisotropies, the renormalised parameter depends quadratically on the bare parameter. Such a behaviour is expected from mean field arguments. However, the corresponding prefactor differs significantly ($\approx 25\%$) from that of the mean field relation \cref{eq:mf}. This introduces a significant systematic error in any numerical study of strongly-coupled lattice QCD.

We find that the whole range of measurements is well described by a simple, one-parameter rational Ansatz (see \cref{fig:gamma-np}b):
\eq{
\frac{\xi(\gamma)}{\gamma^2}  &\approx \kappa + \frac{1}{1 + \lambda \gamma^4}
\label{eq:np}
}
where $\kappa$ is a constant, and $\lambda \mustbe \kappa/(1-\kappa)$, from the requirement that $\xi(1) \mustbe 1$. The approach to the continuous time limit is better captured by Taylor expanding \cref{eq:np} to quadratic order in $1/\xi^2$ (see \cref{fig:gamma-np}c):
\eq{
\frac{\xi(\gamma)}{\gamma^2} &\approx
\kappa \of{1 + \frac{c_1}{\xi^2} + \frac{c_2}{\xi^4}}
\label{eq:np2}
}

The fitted values of $\kappa$ using the Ansatz \cref{eq:np2} -- consistent with those obtained using the Ansatz \cref{eq:np} -- are:
\eq{
\kappa = 
\begin{cases}
0.7795(4), 
& {\rm U(3)}\\
0.7810(8), 
& {\rm SU(3)}
\end{cases}
}
where errors are statistical only. This prefactor is significantly different from the mean field value 1.

Values for U(3) and SU(3) are statistically consistent with each other. This is to be expected: in the continuous time limit, baryons become increasingly static, and their effect on pion currents vanishes at $T=0$.

The Ansatz \cref{eq:np} is also consistent, after differentiation, with the Monte Carlo data for the running anisotropy. In particular, for the isotropic case, instead of the mean field value: \eq{\at{\frac{1}{\xi_{\rm mf}}\der{\gamma}{\xi_{\rm mf}}}{\gamma=1} = 2} we find nonperturbative corrections consistent with:
\eq{
\at{\frac{1}{\xi}\der{\gamma}{\xi}}{\gamma=1} 
\approx 2 + 4\kappa(\kappa - 1)
}

\section{Applications}

In this Section, we use the nonperturbative relation between $\xi$ and $\gamma$, determined above, in order to control the convergence of several physical quantities to their continuous time limits.

First, we examine the $N_t$-dependence of the phase boundary of the $(\upmu_q,T)$ phase diagram of massless SU(3) lattice QCD, and its sensitivity to the anisotropy prescription. Then, we estimate the continuous time values of the static baryon mass $a m_B$, the pion decay constant $aF_\pi$, and the infinite-volume chiral condensate $a^3\Sigma$, in massless U(3) or SU(3) lattice QCD. We use a quadratic Ansatz in $1/\xi^2$, consistent with $O(a^2)$ discretization errors of staggered fermions, to model the anisotropy corrections to the continuous time limit:
\eq{
{\cal O} \approx {\cal O}^{\rm CT}\of{1 + \frac{c_1}{\xi^2} + \frac{c_2}{\xi^4}}
\label{eq:aniso-corrections}
}
where ${\cal O}$ is one of the physical quantities listed above, and ${\cal O}^{\rm CT}$ is the corresponding continuous time value.

For the computation of the pion decay constant and of the chiral condensate, we use the fact that U(3) and SU(3) lattice QCD with massless staggered fermions have an exact $O(2)$ chiral symmetry. At $T=0$ this symmetry is spontaneously broken, and the dynamics of the resulting Goldstone degrees of freedom (pions) are well described by an O(2) sigma model in $d=4$ dimensions. From a finite-size scaling analysis of the discrete O(2) model, it is then possible to extract low-energy quantities like $F_\pi$ and $\Sigma$.

For example, the pion decay constant at $T=0$ can be shown to be related to the helicity modulus $\Upsilon$ \cite{Hasenfratz:1989pk}:
\eq{
a^2 F_\pi^2 = \lim_{N_s\to\infty} a^2 \Upsilon
\label{eq:f-pi}
}
which corresponds, in the diagrammatic representation, to the variance of the conserved pion charges $Q_\mu$ on a hypercubic lattice \cite{Chandrasekharan:2003im}:
\eq{
a^2 \Upsilon = \frac{1}{N_s^2} \vev{Q^2}_{\gamma_{\rm np}}
\label{eq:helicity}
}

In turn, the chiral condensate $\Sigma$ at $T=0$ can be estimated from the finite-size scaling of the chiral susceptibility $\chi$, evaluated on hypercubic lattices. This has been done in $d=3+1$ at finite temperature \cite{Chandrasekharan:2003im}. In our case where $T=0$, chiral perturbation theory of the O(2) model predicts the leading finite-size corrections to be of the form \cite{Hasenfratz:1989pk}:
\eq{
a^6 \chi \approx 
\half a^6 \Sigma^2 N_s^4 
\of{1 + \frac{\beta_1}{a^2 F_\pi^2 N_s^2} 
+ \frac{\alpha}{2 a^4 F_\pi^4 N_s^4} 
}
\label{eq:ChPT}
}
where $\beta_1 = 0.140461$ and $\alpha$ is given by:
\eq{
\alpha = \beta_1^2 + \beta_2 + \frac{1}{8\pi^2} \log\frac{a\Lambda_\Sigma^2 N_s}{\Lambda_M}
}
with $\beta_2 = -0.020305$, and $\Lambda_\Sigma, \Lambda_M$ are renormalisation group invariant scales.
The average value of the chiral susceptibility is estimated using intermediate configurations -- generated with the directed path algorithm -- which sample the mesonic two-point function, as described in \cite{worm}.

\subsection{Phase diagram}

An example of a study that is sensitive to the choice of an anisotropy prescription is the mapping of the phase diagram of massless SU(3) lattice QCD, in the strong coupling limit \cite{deForcrand:2009dh}.

\begin{figure}[t]
\centering
\includegraphics[scale=0.33]{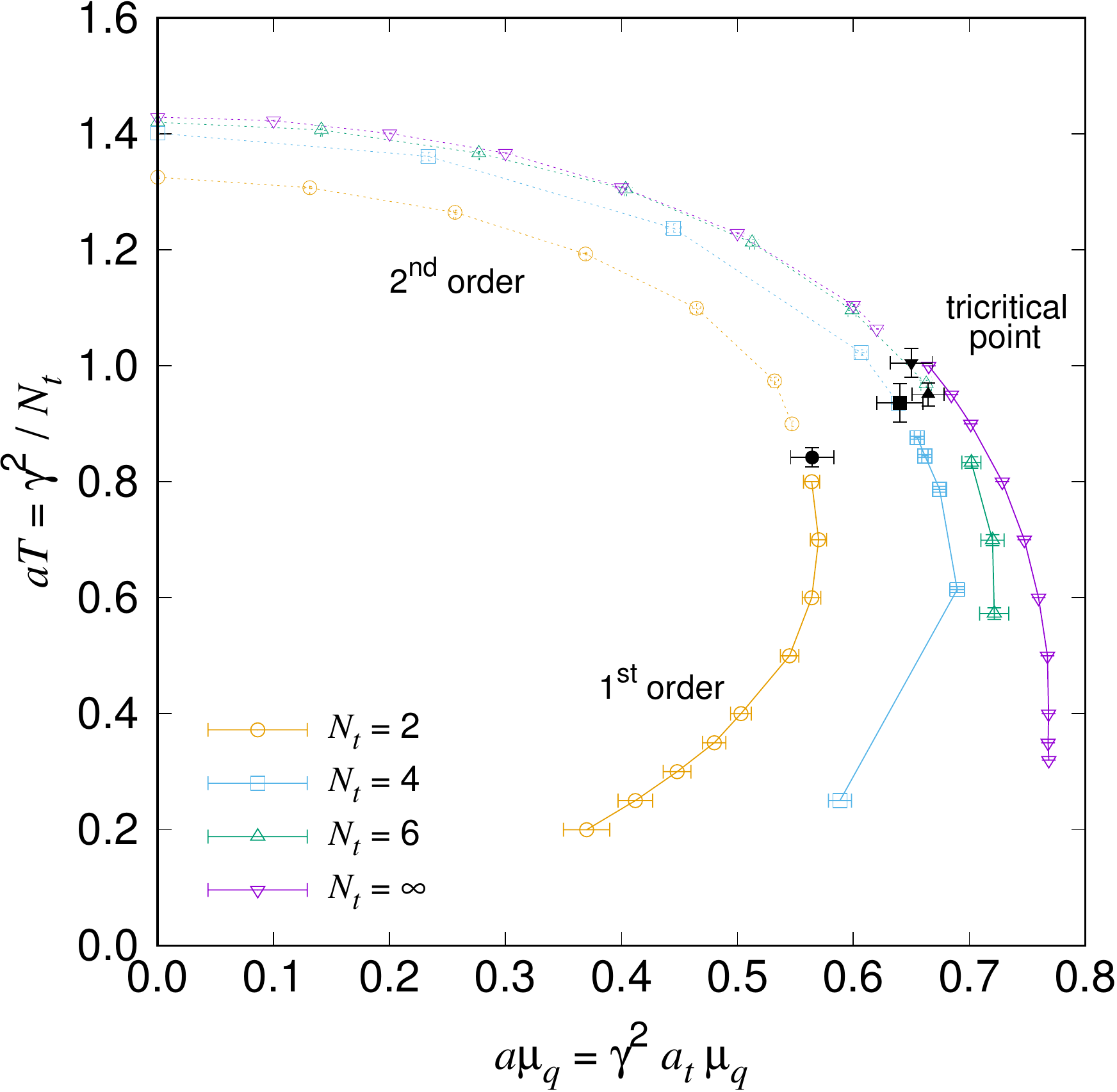}
\vskip 4mm
\includegraphics[scale=0.33]{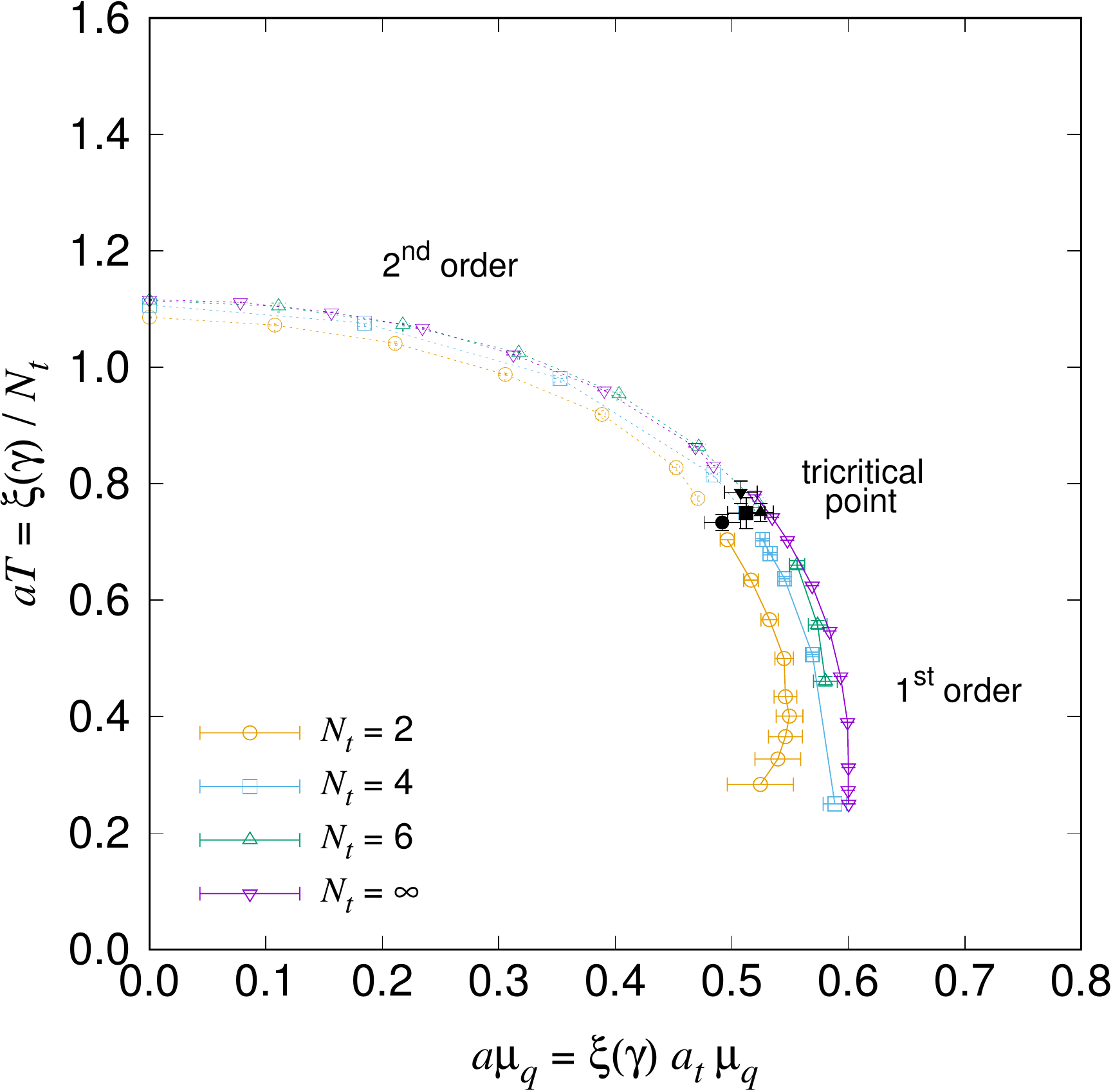}
\caption{
Phase diagram of SU(3) lattice QCD with massless staggered fermions, in the strong coupling limit, in which the anisotropy is set using mean field (top) \cite{deForcrand:2014tha}, or using the present nonperturbative prescription (bottom). Under the nonperturbative prescription \cref{eq:np}, the $N_t$-dependence of the phase boundary and of the tricritical point decreases substantially. Also, the tricritical couplings on the horizontal and vertical axes both decrease by $\approx 25\%$. The $N_t=\infty$ data is produced from simulations directly in the continuous time limit \cite{Unger:2011in}.
\label{fig:phase-diagram}}
\end{figure}

The phase boundary separating the chirally broken phase at low $(\upmu_q,T)$ and the chirally symmetric phase at high $(\upmu_q,T)$ is determined by monitoring the chiral condensate $a^3\Sigma$ during Monte Carlo simulations, using directed path algorithms and sign reweighting for importance sampling on moderate volumes (see \cref{fig:phase-diagram}).

For fixed $N_t$, the temperature is varied implicitly through the bare coupling $\gamma$ \cite{deForcrand:2014tha}. 
Assuming the mean field relation \cref{eq:mf}, the observed phase boundary has a strong dependence on $N_t$ (see \cref{fig:phase-diagram}, top), which makes its interpretation questionable. 
This systematic error is dramatically reduced by using the nonperturbative prescription \cref{eq:np} for the renormalised anisotropy (see \cref{fig:phase-diagram}, bottom). Note that, under the nonperturbative prescription, the tricritical couplings on the temperature and chemical potential axes both decrease by $\approx 25\%$. 

Moreover, analytic studies of the phase diagram generally consider Euclidean time as continuous \cite{KawamotoDamgaard}, and should be compared with the $N_t = \infty$ data only.

\subsection{Static baryon mass}

The static baryon mass $am_B$ is another observable for which the inexact calibration of anisotropy can have a strong effect. This observable can be determined using the ``snake algorithm'' \cite{snake}, which samples partition functions $Z_k$ describing the system with an open baryonic segment of length $k$:
\eq{
a m_B = \frac{\xi}{N_t} \sum_{k=0}^{N_t-2} \log \frac{Z_{k+2}}{Z_k}
\label{eq:baryon-mass-snake}
}

\begin{figure}[t]
\centering
\includegraphics[scale=.33]{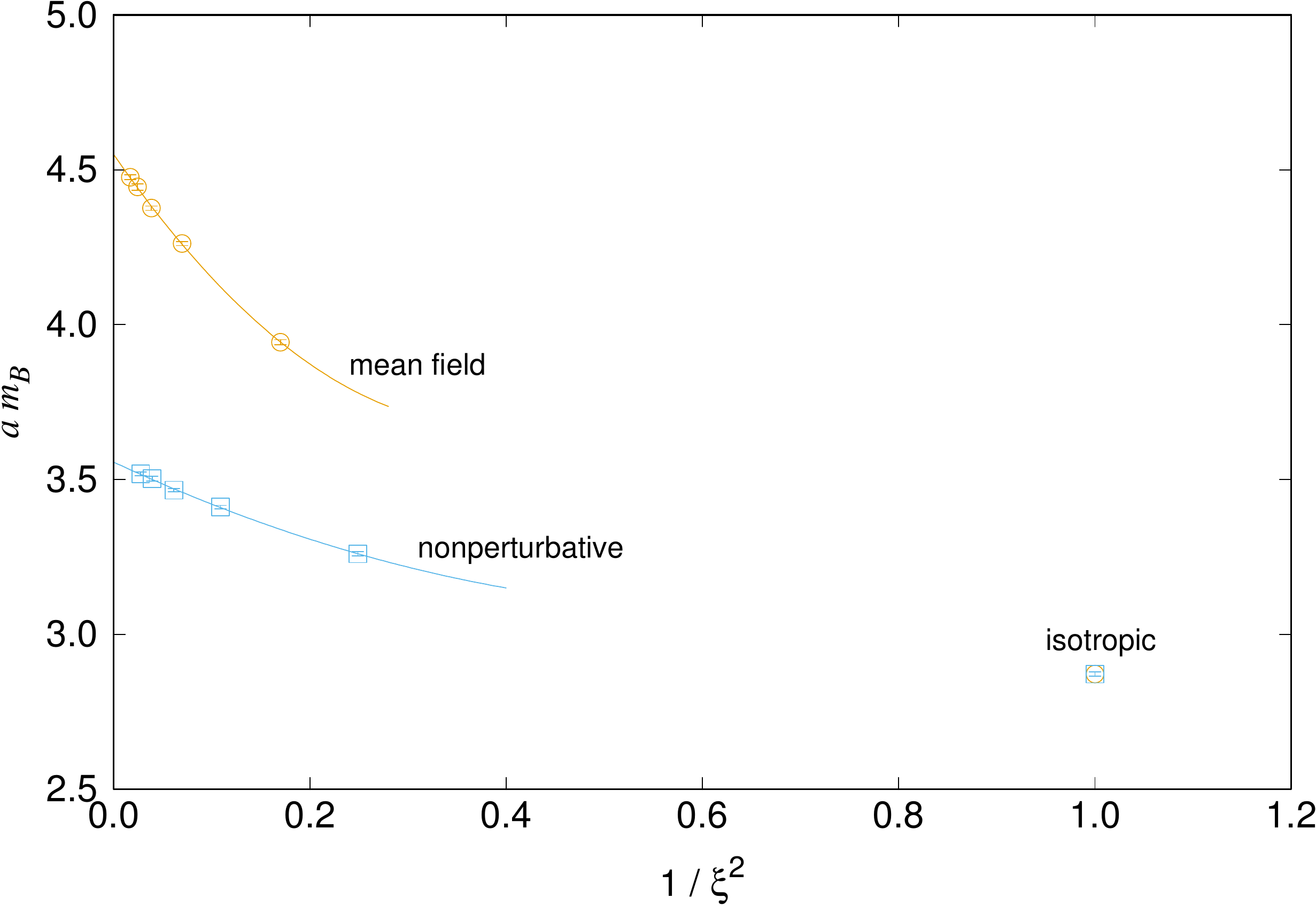}
\caption{
Effect of the physical anisotropy on the static baryon mass, in massless SU(3) lattice QCD. The anisotropy corrections to the continuous time limit ($\xi \to \infty$) are well described by a quadratic Ansatz in $1/\xi^2$. The baryon mass is heavier on anisotropic lattices than on isotropic lattices, where its value is $am_B \approx 2.88$ \cite{deForcrand:2009dh}. With the anisotropy set using mean field, the baryon mass receives an $\approx 50\%$ correction in the continuous time limit with respect to the isotropic case, while under the present nonperturbative prescription it only receives an $\approx 20\%$ correction.
\label{fig:baryon-mass}}
\end{figure}

We simulate massless SU(3) lattice QCD for different anisotropies using the snake algorithm, and estimate $am_B$ as a function of $\xi$ (see \cref{fig:baryon-mass}). Under the two anisotropy prescriptions, \cref{eq:mf,eq:np}, baryon masses differ by $\approx 25\%$ at large $\xi$. In this regime, the fitting Ansatz \cref{eq:aniso-corrections} describes the data well. The vertical intercepts give the values of the static baryon mass in the continuous time (CT) limit:
\eq{
(a m_B)^{\rm CT} =
\begin{cases}
4.550(8), & {\rm mean~field} \\
3.556(6), & {\rm nonperturbative}
\end{cases}
}

On an isotropic lattice, static baryons have mass $am_B \approx 2.88$ \cite{deForcrand:2009dh}, and become heavier with anisotropy. In the continuous time limit, the baryon mass is only $\approx 20\%$ heavier than the isotropic case, when using the nonperturbative prescription for the anisotropy, as compared with the $\approx 50\%$ difference when using mean field. 

\subsection{Pion decay constant}

\begin{figure}[t]
\centering
\includegraphics[scale=.33]{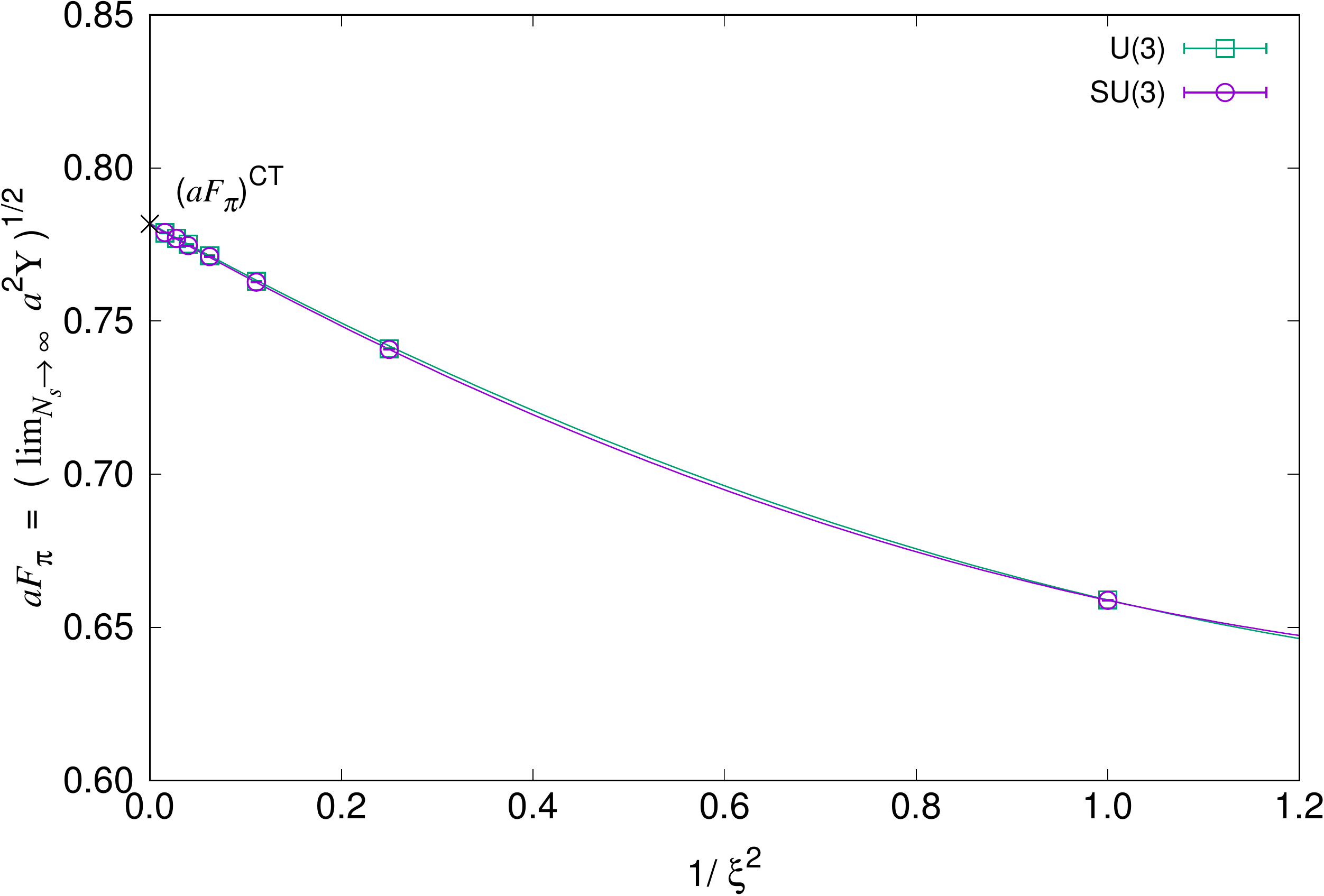}
\vskip 4mm
\includegraphics[scale=.33]{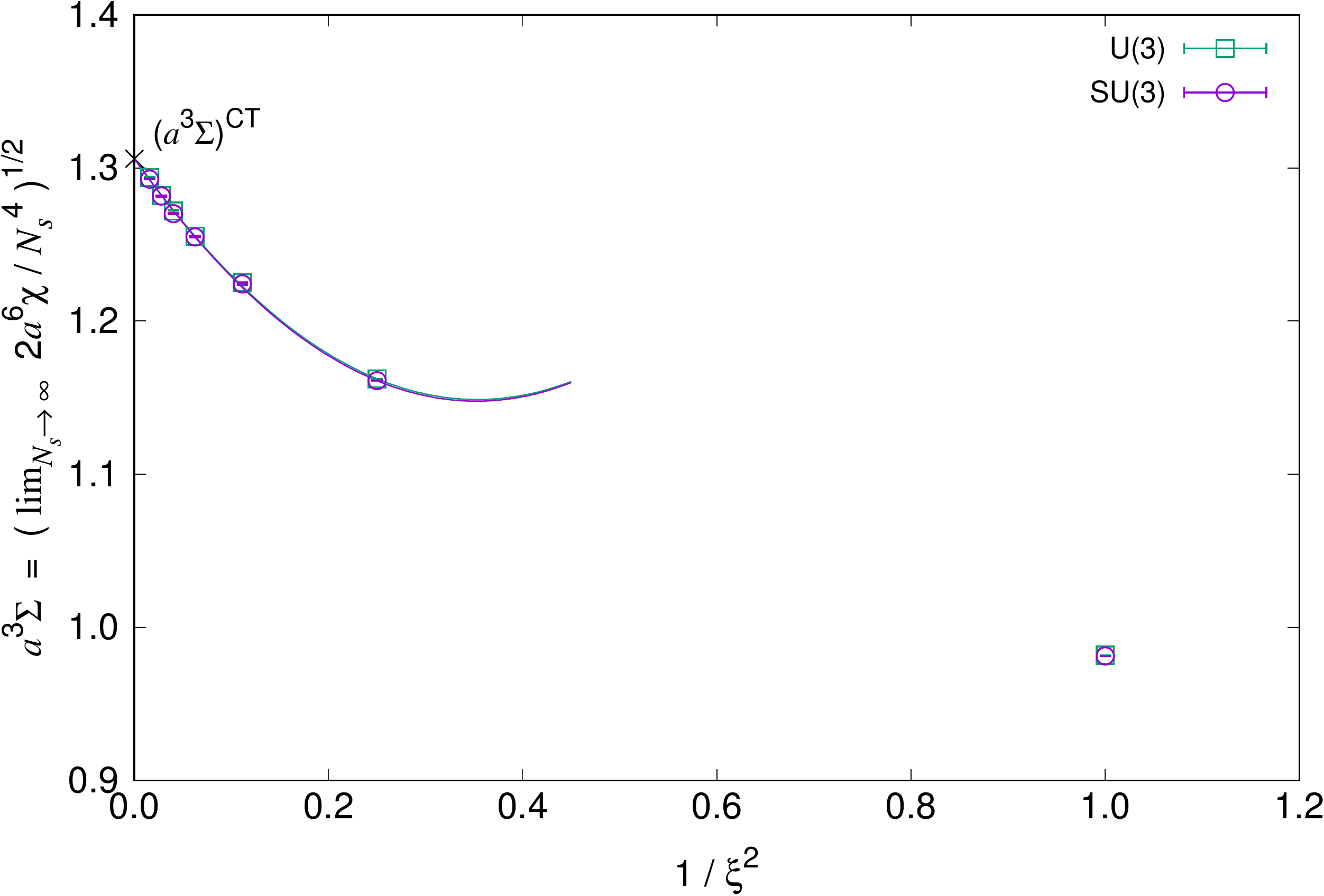}
\caption{Effect of the physical anisotropy on the pion decay constant (top) and on the chiral condensate (bottom), in massless U(3) and SU(3) lattice QCD. The anisotropy corrections to the continuous time limit ($\xi \to \infty$) are rather small, and well described by a quadratic Ansatz in $1/\xi^2$. The baryonic corrections to the U(3) helicity modulus are negligible. In both graphs, the isotropic points $\xi = 1$ are not included in the quadratic fits: the intersection of the fitting curves with the isotropic point in the top graph is accidental.
\label{fig:helicity_cc}}
\end{figure}

Using our nonperturbative prescription for $\xi$, we can obtain reliable estimates of several physical quantities in the continuous time limit, \eg the pion decay constant, $a F_\pi$. In order to estimate this quantity, we measure the helicity modulus \cref{eq:helicity} for several finite hypercubic lattices and values of $\xi$. The results are summarised in \cref{tab:gamma-u3,tab:gamma-su3}, and displayed in \cref{fig:helicity_cc} (top). The pion decay constant (squared) corresponds to the thermodynamic limit of the helicity modulus, in accordance with \cref{eq:f-pi}.

Again, the numerical data can be suitably fitted using the Ansatz \cref{eq:aniso-corrections}. At large $\xi$, the anisotropy corrections are rather small. The vertical intercepts give the values of the pion decay constant in the continuous time limit at $T=0$: \footnote{New, direct measurements of $a F_\pi$ in the continuous time limit \cite{WolfgangMarc} are consistent with our extrapolation.}
\eq{
(a F_\pi)^{\rm CT} = 
\begin{cases}
0.7820(2), & {\rm U(3)}\\
0.78171(4), & {\rm SU(3)}
\end{cases}
\label{eq:fpi-values}
}
Again, U(3) and SU(3) are equivalent in the thermodynamic and continuous time limits, within errors.

\subsection{Chiral condensate}

We also estimate accurate values for the infinite-volume chiral condensate $a^3 \Sigma$, by analysing the finite-size scaling of the chiral susceptibility $a^6 \chi$, using chiral perturbation theory, and by using our nonperturbative prescription for the lattice anisotropy.

To this end, we estimate the chiral susceptibility density $a^6\chi/N_s^4$ (as in \cite{worm}) for several finite hypercubic lattices and values of $\xi$ (see \cref{tab:gamma-u3,tab:gamma-su3}). We estimate $a^6 \Sigma^2$ at finite $\xi$ by extrapolating $a^6\chi/N_s^4$ to the thermodynamic limit, modelling the finite-size corrections in accordance with chiral perturbation theory, see \cref{eq:ChPT}.

The dependence of $a^3\Sigma$ on $\xi$ is again well described by the Ansatz \cref{eq:aniso-corrections} (see \cref{fig:helicity_cc}, bottom). The vertical intercepts give the values of the chiral condensate in the continuous time limit at $T=0$:
\eq{
(a^3 \Sigma)^{\rm CT} =
\begin{cases}
1.3063(9), & {\rm U(3)}\\
1.306(1), & {\rm SU(3)}
\end{cases}
}
As before, U(3) and SU(3) are equivalent in the thermodynamic and continuous time limits, within errors. We also observe that, when keeping $\beta_1$ as a free parameter in \cref{eq:ChPT}, the finite-size fits are consistent with its theoretical value.

\section*{Conclusion\label{sect:conclusion}}

It is very important to have a precise scale for the lattice anisotropy. Even though mean field captures the correct power scaling of the renormalised anisotropy for asymptotically large values of the bare anisotropy, namely $\xi \sim \gamma^2$, it fails to predict the nonperturbative prefactor. The discrepancy between the mean field and nonperturbative prefactors introduces systematic errors of the same magnitude in many physical quantities of interest, particularly in the continuous time limit.
This should be kept in mind when comparing strong-coupling Monte Carlo results and analytic mean field results, since the latter are usually formulated in continuous time.

In the dimer representation of the strong coupling limit of lattice QCD with massless staggered fermions, we have proposed a simple method to determine the nonperturbative dependence $\xi(\gamma)$ between the bare and renormalised anisotropy couplings. The method is amenable to Monte Carlo simulations using very efficient directed path algorithms which, together with the multi-histogram reweighting method, allows us to determine $\xi(\gamma)$ with high precision. In the end, the nonperturbative prefactor is observed to be off by $\approx 25\%$ with respect to the mean field prefactor.

As an application, we revisit the phase diagram of SU(3) lattice QCD \cite{deForcrand:2009dh}, and update it using our nonperturbative relation $\xi(\gamma)$. A strong dependence of the phase boundary on $N_t$, introduced by the mean field anisotropy, essentially vanishes. The new locations of the phase boundary and of the tricritical point reveal corrections of $\approx 25\%$, in the chemical potential and temperature, compared with the old mean field values. We also compute the mass of the static baryon in the continuous time limit, which again receives corrections of $\approx 25\%$ compared with the mean field value. These corrections are the direct consequence of the $\approx 25\%$ correction to the mean field prefactor to $\xi(\gamma)$ mentioned above.

We also estimate the values of the pion decay constant, $a F_\pi$, and of the infinite-volume chiral condensate, $a^3\Sigma$, in massless lattice QCD in the strong coupling limit at $T=0$. The anisotropy corrections to these quantities are small, and provide a reliable extrapolation to their continuous time limits. 

Even though the strong coupling limit of lattice QCD is unphysical, it may still be of interest to compare its predictions with those of continuum QCD, in the regime where  chiral symmetry is spontaneously broken. For example, the strong-coupling SU(3) lattice value of the pion decay constant \cref{eq:fpi-values}, in units of the critical temperature $aT_c \approx 1.089$ \cite{deForcrand:2017obe}, is $F_\pi/T_c \approx 0.72$, which is about 15\% above the continuum QCD value.

Our approach can be generalized to the case of massive quarks. As before, in a hypercubic box the variances of the spatial and the temporal pion charges \cref{eq:pion-charge} can be required to be equal. Since they still scale as in \cref{eq:charge-scaling-t,eq:charge-scaling-s}, the renormalisation criterion \cref{eq:renormalization-criterion} is justified. What changes is that the pion charges are no longer conserved as per \cref{eq:pion-current-conservation}, \ie have different values on parallel codim-1 lattice slices. A sensible observable is the average over such parallel slices of the variance of the pion charge on each slice. Thus, the setting of the anisotropy should be performed in a fixed volume $L^4$, characterised by the value of $m_\pi L$. This implies a fine-tuning of the quark mass, in order to keep fixed $m_\pi L$ while the bare anisotropy $\gamma$ is varied. Alternatively, the anisotropy may also be set by keeping $\xi$ and $m_q L$ = $N_t a_t m_q$ fixed while varying $\gamma$ \cite{WolfgangMarc}.

It may also be possible to extend the present study to finite $\beta$, in the framework of the $O(\beta)$ partition function defined in \cite{deForcrand:2014tha}. The new occupation numbers (associated with plaquettes) introduce new Grassmann constraints on the extended configuration space. Such constraints may be used to construct analogues of the pion current, which would include plaquette corrections. In the chiral limit, we expect such currents to be conserved. The associated conserved charges could then be used to define nonperturbative renormalisation criteria for the (independent) spatial and temporal gauge couplings. An extension of this program to finite quark mass would be similar to the above proposal for $\beta=0$.

\bigskip
\bigskip
\section*{Acknowledgements}
We are very grateful to Oscar \AA kerlund, Tobias Rindlisbacher, and Paul Romatschke for many useful discussions. 
We are also grateful to the Mainz Institute for Theoretical Physics (MITP) for its hospitality and its partial support during the completion of this work. 
Numerical simulations were performed on the Brutus and Euler clusters at ETH Zürich, 
and on the OCuLUS cluster at PC$^2$ (Universit\"{a}t Paderborn).
This work is supported by the Swiss National Science Foundation under the grant 200020\_162515. W. U. acknowledges support by the Deutsche Forschungsgemeinschaft (DFG) through the Emmy Noether Program under grant No. UN 370/1 and through the grant CRC-TR 211 ``Strong-interaction matter under extreme conditions.''


\end{document}